\documentclass[10pt,journal,compsoc]{IEEEtran}

\usepackage{booktabs} 
\usepackage{siunitx}
\usepackage{amsmath,xcolor,pifont}

\usepackage[colorinlistoftodos]{todonotes}

\usepackage{array}
\newcolumntype{L}[1]{>{\raggedright\let\newline\\\arraybackslash\hspace{0pt}}m{#1}}
\newcolumntype{C}[1]{>{\centering\let\newline\\\arraybackslash\hspace{0pt}}m{#1}}
\newcolumntype{R}[1]{>{\raggedleft\let\newline\\\arraybackslash\hspace{0pt}}m{#1}}

\usepackage{color}
\usepackage{multirow}
\usepackage{makecell}
\usepackage{tabularx}
\usepackage{listings}
\usepackage{graphicx}
\usepackage{caption}
\usepackage{subcaption}

\definecolor{mygreen}{rgb}{0,0.6,0}
\definecolor{mygray}{rgb}{0.5,0.5,0.5}
\definecolor{mymauve}{rgb}{0.58,0,0.82}
\definecolor{pblue}{rgb}{0.13,0.13,1}
\definecolor{pgreen}{rgb}{0,0.5,0}
\definecolor{pred}{rgb}{0.9,0,0}
\definecolor{pgrey}{rgb}{0.46,0.45,0.48}

\usepackage[ruled]{algorithm2e} 

\SetAlFnt{\small}
\SetAlCapFnt{\small}
\SetAlCapNameFnt{\small}
\SetAlCapHSkip{0pt}
\IncMargin{-\parindent}

\begin{document}
\newtheorem{defn}{Definition}


\title{Automated Extraction of Personal Knowledge from Smartphone Push Notifications}

\author{Yuanchun~Li,
        Ziyue~Yang,
        Yao~Guo,
        Xiangqun~Chen,
        Yuvraj~Agarwal,
        Jason~I.~Hong
\IEEEcompsocitemizethanks{\IEEEcompsocthanksitem Y. Li, Z. Yang, Y. Guo, and X. Chen are with the Key Laboratory of High Confidence Software Technologies (Peking University), Ministry of Education, Beijing 100871, China.\protect\\
E-mail: {liyuanchun, ziyue.yang, yaoguo, cherry}@pku.edu.cn
\IEEEcompsocthanksitem Y. Agarwal and J. Hong are with the School of Computer Science, Carnegie Mellon University, PA 15213, USA.\protect\\
E-mail: {yuvraj.agarwal, jasonh}@cs.cmu.edu}
\thanks{Manuscript received May 20, 2018.}}

\markboth{Technical Report}%
{Li \MakeLowercase{\textit{et al.}}: Automated Extraction of Personal Knowledge from Smartphone Push Notifications}

\IEEEtitleabstractindextext{%
\begin{abstract}
Personalized services are in need of a rich and powerful personal knowledge base, i.e. a knowledge base containing information about the user. This paper proposes an approach to extracting personal knowledge from smartphone push notifications, which are used by mobile systems and apps to inform users of a rich range of information. Our solution is based on the insight that most notifications are formatted using templates, while knowledge entities can be usually found within the parameters to the templates. As defining all the notification templates and their semantic rules are impractical due to the huge number of notification templates used by potentially millions of apps, we propose an automated approach for personal knowledge extraction from push notifications. We first discover notification templates through pattern mining, then use machine learning to understand the template semantics. Based on the templates and their semantics, we are able to translate notification text into knowledge facts automatically. Users' privacy is preserved as we only need to upload the templates to the server for model training, which do not contain any personal information. According to our  experiments with about 120 million push notifications from 100,000 smartphone users, our system is able to extract personal knowledge accurately and efficiently.
\end{abstract}




}



\maketitle

\IEEEdisplaynontitleabstractindextext
\IEEEpeerreviewmaketitle
\IEEEraisesectionheading{\section{Introduction}\label{sec:introduction}}


\IEEEPARstart{P}{ush} notifications are widely used on mobile devices such as iPhone or Android smartphones. A push notification is a message that pops up on a mobile device and can be used for multiple purposes, such as SMS or social networking updates (e.g. your friend Alice sent you a message), travel schedule changes (e.g. your flight to Pittsburgh is canceled), and shopping order delivery messages (e.g. the clothes you purchased has been shipped), just to name a few.
Each user receives about 63.5 notifications per day on his/her smartphone \cite{noti_in_situ_study}.


This paper proposes an approach to extracting personal knowledge (\texttt{<user, relation, entity>} triples) from smartphone push notifications.
Personal knowledge is a structured form of data that contains information about users' profile, behaviors, interests, etc. Such knowledge is important and useful for various mobile applications and mobile services, such as recommender systems \cite{targeting_ad,TMC:Wang:Friendbook}, virtual personal assistants \cite{gruber2009siri}, and authentication systems \cite{TMC:Guo:Authentication}.
Researchers have also proposed methods for extracting personal knowledge from various other kinds of data sources such as user utterances \cite{personalKnowledgeUtterances} and communication logs \cite{use_case6:cscw2013:social_relationships}.
In fact, many companies, especially smartphone vendors, have already started to make use of the personal knowledge extracted from different sources on smartphones to provide better services (e.g., emails, SMS messages and calendars).
For example, Google\footnote{https://developers.google.com/gmail/markup/google-now} extracts and summarizes flight and hotel reservation information from emails with markup \cite{schema.org}.
Apple Siri\footnote{https://www.apple.com/ios/siri/} reads users' calendar events to answer questions like ``when is my next appointment''.
However, these approaches only deal with specific categories of personal knowledge, where the information is well-structured or programmatically available.
Compared to them, push notifications are a more natural source of personal knowledge as they act like a proxy to many other data sources.



Extracting personal knowledge in general on smartphones is a difficult task.
On one hand, there exists abundant personal information on smartphones, which can be exploited for many apps to provide better services to end users. On the other hand, it is not desirable to obtain full permissions to access personal information directly as protecting user privacy has also become a first-order priority for mobile apps \cite{Wiese:evolving}.

In contrast, using push notifications as sources for personal knowledge offers several key benefits.
First, push notifications contain and summarize a rich range of important personal information, such as user profiles, social relationships and information on everyday life.
Second, push notifications are well-structured, as most of them are generated automatically using fixed templates, thus simplifying the task of extracting useful information from them.
Third, notifications offer a uniform way of accessing data siloed across many apps.

Similar to the general knowledge base population (KBP), extracting personal knowledge from push notifications can be viewed as a slot filling task \cite{Ji:2011:KBP}.
The entities related to the user (e.g. the Twitter accounts that the user follows, the products that the user purchases, etc.) are reserved as slots, and the goal of personal knowledge extraction is to collect entity values (slot fillers) from the large-scale push notifications.
The state-of-the-art approaches \cite{zhang2016joint,yao2014recurrent} for slot filling model the problem as a sequence labeling task and use RNNs to find both the boundaries and labels of slot fillers.
However, the entity values of personal knowledge in push notifications are often arbitrary phrases (e.g. @realDonaldTrump, iPhone X 64G Silver, etc.), making it extremely hard to find the entity boundary through sequence labeling.

A more straightforward solution is to define a template for each kind of push notifications manually, one that captures the semantics embodied in the notification. Once we have these templates, it becomes very easy to identify the relevant notifications and extract the related components to form a database of personal knowledge about the user. 
For example, here is a typical notification template: ``Dear \texttt{\$param1}, here are some \texttt{\$param2} job opportunities for you''. Besides the structure, we can easily understand that \texttt{\$param1} is the name of the user, \texttt{\$param2} is the user's profession, and the user is hunting for jobs. Applying this rule to a specific push notification ``Dear David, here are some software engineer job opportunities for you'', we are able to extract the parameters (\texttt{David} and \texttt{software engineer}) and generate knowledge triples: \texttt{<user, name, David>}, \texttt{<user, profession, software engineer>}, and \texttt{<user, status, job\_hunting>}.

However, 
the above mentioned solution require defining the patterns or templates manually, and as such does not scale well, especially as the number of apps increases. Each of these apps might use a different template to construct their notifications. 
To solve this challenge, this paper proposes an automated approach to identify notification templates and to learn their semantics. 
Specifically, our proposed approach includes the following steps: it first discovers the notification templates on the device, then uploads the templates to the server for offline learning to train a model to understand the semantic meaning of each template, and finally, it is able to extract personal knowledge based on the server-trained model.

We achieve the following goals with the proposed approach: (1) we are able to automatically identify the templates for different types of notifications, including formerly unseen new templates; (2) we can also understand the meaning of each new template through an offline learning phase; (3) because we only need the templates (without specific user information) to train the model, we do not need to send sensitive user information out of the devices, thus helping preserve user privacy during the whole process.
To evaluate our approach, we conducted experiments on around 120 million real notifications from 100,000 smartphone users.
The results show that our system is able to discover notification templates with a precision of 86.8\% and understand the semantics of unseen templates accurately (around 83\% F1-score for templates of new apps and 91\% F1-score for new templates of existing apps).
We also demonstrate that the discovered templates and the semantic model can be directly used to extract personal knowledge from push notifications.

This paper makes the following main contributions:
\begin{itemize}
\item To the best of our knowledge, this is the first work to propose that push notifications can be used as a data source for personal knowledge extraction on mobile devices such as smartphones. We also introduce an automated and privacy-preserving method to extract personal knowledge from push notifications, based on the insight that notification templates can be learned offline and used locally on one's device to extract personal knowledge facts.
\item We implement a prototype system for personal knowledge extraction on Android. The system can run on smartphones to support personalized services such as recommender systems and virtual personal assistants.
\item We evaluate our approach on around 120 million real-world push notifications from 100,000 users. The results show that there are around 5.7\% of notifications containing various types of personal knowledge, and our method is able to extract these personal knowledge with high accuracy. 
\end{itemize}

\section{Background}

\subsection{Push Notifications}

Push notifications serve as a core feature for mobile devices such as smartphones and tablets. 
They are mainly used by the operating system and smartphone apps to inform users of various of events, such as the availability of a software update, the arrival of a message, the status update of an online purchase, the recommendation of news and articles, etc.
As notifications can be displayed without activating the apps' normal UI, they are a preferable way used by app developers to deliver information to users promptly. As a result, push notifications sometimes contain valuable information.

The content of smartphone push notifications can be generated either locally or remotely. Most operating systems provide APIs for apps to display notifications locally. For example, Android allows apps to define a \texttt{Notification} instance, set a title and a text body, and send it as an \texttt{Intent} to display. Most notifications of system events such as alarms and device status updates are generated with this method.
Many systems and third-party services also provide a way for developers to construct notifications on the server, then push them to the client devices. Remotely-generated push notifications are more dynamic and less structured as compared to local notifications, since any online-service notifications such as messages, news, and advertisements can be generated remotely.

Most push notifications are automatically generated with templates. However, because each push notification typically contains personalized content, an app can customize the template parameter for each user to achieve this goal, while the templates remain the same across different users. Thus, it is possible to extract the personal information from push notifications once we know the templates.

\subsection{Knowledge representation}

\begin{table*}[tbp]
\centering
\caption{The ontology of the personal knowledge considered in this paper. We considered 11 types of knowledge relations (column 2) in 4 categories (column 1). For each relation, we show several examples of entities (column 3) and a sample notification text (column 4). Note that the examples are simplified (and translated) for better presentation.}
\label{table:ontology}
\begin{tabular}{clll}
\hline
\textbf{Category}                    & \textbf{Relation} & \textbf{Example entities}           & \textbf{Example notification}            \\ \hline
\multirow{4}{*}{User profile}        & name       & Alice, Bob1997, ...        & Hi Alice, here are some recommended reads for you.   \\
                                     & gender       & male, female, ...       & Dear Mr. Li, please review your receipt.        \\
                                     & profession    & doctor, software engineering, ...       & 7 software engineer positions for you: ...        \\
                                     & status     & in\_college, job\_hunt, ...       & Facebook: found 9 classmates in Stanford University.              \\ \hline
\multirow{2}{*}{Social} & follows     & Justin Bieber, @realDonaldTrump, etc.     & Justin Bieber posted a new photo.                   \\
                                     & isFriendOf        & Candy's Mother, David, etc.     & David sent you a message: ...         \\ \hline
\multirow{2}{*}{Location}     & livesNear     & Beijing, MIT campus, etc.    & Beijing weather today: 6 C, sunny.           \\
                                     & travelsTo  & Sweden, Tokyo, etc.   & Flight CU1234 from Beijing to Tokyo is going to take off.  \\ \hline
\multirow{3}{*}{Shopping}     & purchases      & iPhone X 64G Silver, milk powder, etc.    & Your order iPhone X 64G Silver has been shipped.           \\
                                     & wantsToBuy  & NIKE Men's Roshe Run Size 10, beer, etc.   & The beers in your shopping cart is on sale.  \\
                                     & visitsMerchant     & Walmart, Wendy's, etc.   & Thank you for shopping at Walmart. \\ \hline
\end{tabular}
\end{table*}

Existing knowledge bases such as YAGO \cite{yago}, Freebase \cite{freebase} and Google Knowledge Graph \cite{knowledgeVault} use relational knowledge representations. Information is modeled in the form of entities and relations between them.
Such kind of representation has been widely used in the area of logic and artificial intelligence \cite{davis1993knowledge}.

The W3C Resource Description Framework (RDF) \cite{www2014rdf} defines an abstract syntax for relational information representation. The core structure of the abstract syntax is a set of \emph{subject, predicate, object} (SPO) triples, where subjects and objects are entities, while predicates are defined as the relations between them. All existing knowledge bases can be represented with such SPO triples.

Similar to the world's general knowledge bases, the facts in personal knowledge bases can also be represented as SPO triples.
Li \textit{et al.} \cite{personalKnowledgeUtterances} represent personal knowledge as a user-centered graph, in which the subjects of all knowledge triples are the user. They follow the Freebase semantic knowledge graph schema, including 18 types of \texttt{<user, relation, entity>} triples, such as \texttt{<user, place\_of\_birth, New York City>}, \texttt{<user, parents, Rosa>}, etc.
We follow the definition of Li \textit{et al.}, but we use a different set of relations that frequently appear in push notifications.

The ontology of the personal knowledge considered in this paper is shown in Table~\ref{table:ontology}.
we consider four common categories of personal knowledge, including user profile, social relationship, location and shopping. Other categories of knowledge can be easily added in the future.

\section{Personal Knowledge Extraction from Push Notifications}

We propose an automated approach to extract personal knowledge facts from push notifications on smartphones. The problem is defined as follows. Suppose there are a set of smartphone users, each with a list of push notifications. Our goal is to extract knowledge triples (\texttt{<user, relation, entity>}) from the notification content for each user, with little-to-none manual efforts.

Of course, developers can always define templates manually. However, because there are too many notifications, it takes a lot of efforts to manually identify and define all the different templates for all apps. Moreover, there are always new templates, which cannot be covered by existing templates defined manually. In contrast, we expect that an automated approach can identify new notification templates automatically, as well as the meaning for each notification.

Our approach is mainly based on the observation that knowledge is formatted into notifications with templates. The templates can then help identify knowledge entities and understand their meanings as well.
We aim to identify the notification templates automatically, and then understand their structure and meanings through machine learning.

\begin{figure*}[tbp]
\centering
\includegraphics[width=7in]{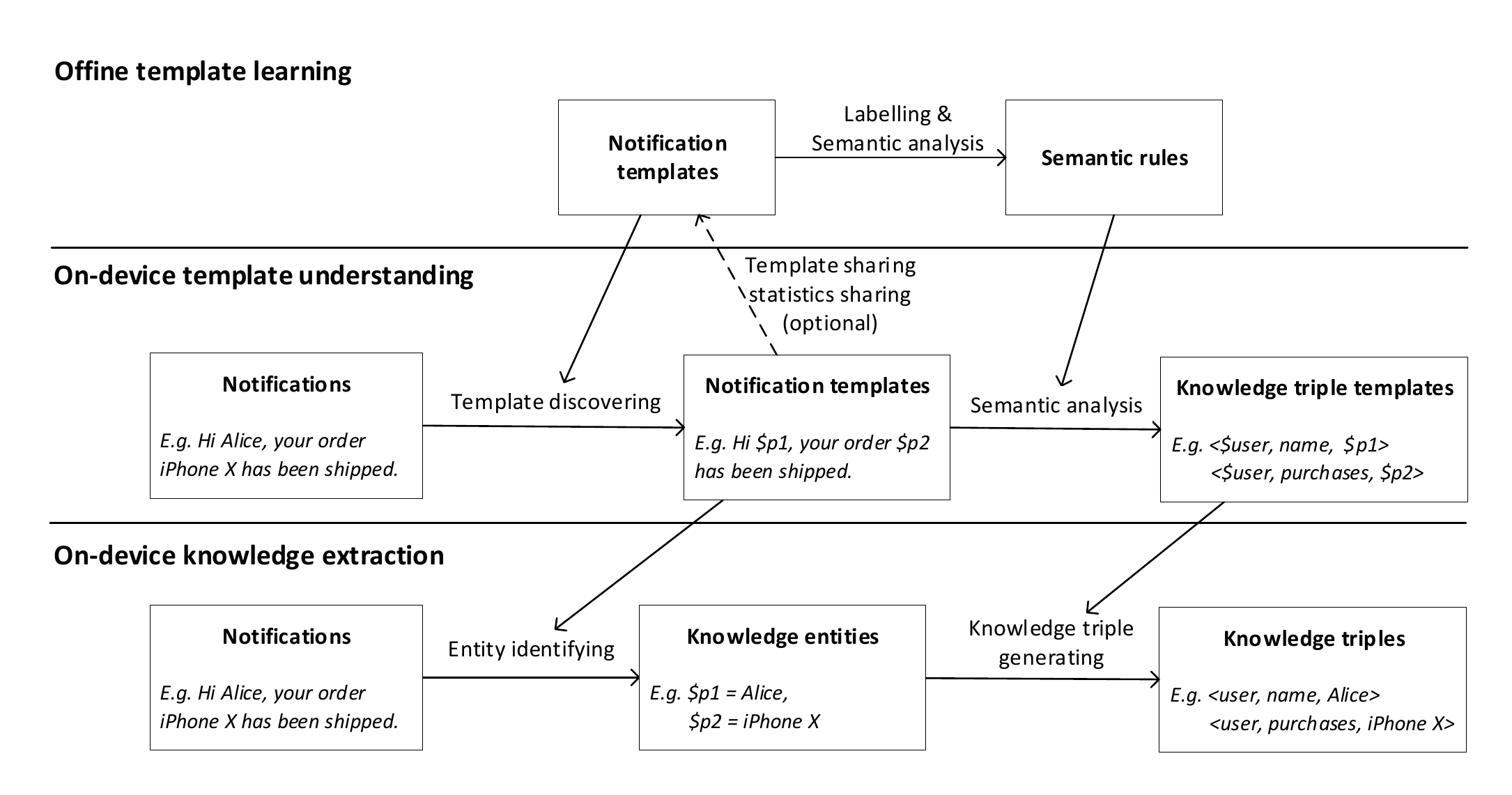}
\caption{An overview of our approach. We first identify templates from the notifications on each user device, then infer template semantic rules using a server-trained semantic model. The templates are used to identify personal knowledge entities, while the template semantic rules help generate knowledge triples.}
\label{figure:overview}
\end{figure*}

\subsection{Approach Overview}

Figure~\ref{figure:overview} shows an overview of our proposed approach in extracting personal knowledge from push notifications.
The approach consists of three phases: template learning, template understanding, and knowledge extraction.
The template learning phase runs on the server and the other two phases run on users' devices.
The main purpose of template learning is to train a machine learning model to understand the semantics of notification templates.
Then in the template understanding phase, the trained model can be used to infer what types of personal knowledge triples that each notification template may express.
The discovered templates and the inferred semantics are then used to identify template parameters (i.e. entities) from notification text and generate personal knowledge triples.

Consider an example notification with order shipping information: ``Your order iPhone X has been shipped''.
For each type of notification, we assume that there are other similar notifications on the device, such as ``Your order Nike Running Shoes has been shipped''.
By mining patterns from all notifications, we can discover the template for this notification: ``Your order \texttt{\$param} has been shipped'', where \texttt{\$param} is a parameter to the template.
The template parameters are potentially personal knowledge entities, as they are usually customized for different users.

However, discovering the template of a notification is only the first step. Once we extract the relevant elements from a notification, we still need to understand the meaning of each component from the notification. 
In order to infer whether the template is a personal knowledge template and what knowledge triples the template may have, we use a server-trained semantic model to understand the template.
The semantic analysis is modeled as a multi-label classification problem: given a notification template, predict what kinds of personal knowledge triples it may express.
Specifically, given the template ``Your order \texttt{\$param} has been shipped'', the semantic model will predict a \texttt{purchases} relation for \texttt{param}, which leads to a knowledge triple templates \texttt{<user, purchases, \$param>}. The model will also try to predict no-parameter knowledge triples (such as \texttt{<user, gender, male>}) based on the whole template.

The mapping from the notification template to the knowledge triple templates is referred to as a \textit{template semantic rule} in this paper.
By applying the template semantic rule to the original notification content, we are able to extract the knowledge triples: \texttt{<user, purchases, iPhone X>}.

The whole process introduces two main challenges: discovering notification templates and understanding template semantics.
The following sections describe how we solve these problems through pattern mining and machine learning, respectively.

\subsection{Template Discovering}
\label{section:template_discovering}


The purpose of template discovering is to recover the templates that are used by app developers to generate notifications.
Many notifications are generated by programs, and typically use templates internally. However, there are two main challenges in discovering the template for an arbitrary notification.
The first challenge is that templates vary quite a bit in terms of personal data used, across different apps, and across different app versions. Altogether, these differences make it impractical, if not impossible, to manually summarize a complete list of templates. The second challenge is that the notifications on users' smartphones are usually privacy-sensitive. Thus uploading all notifications to a server for joint analysis is undesirable because it may cause the leakage of personal information.

\begin{figure}[tbp]
\centering
\includegraphics[width=3.5in]{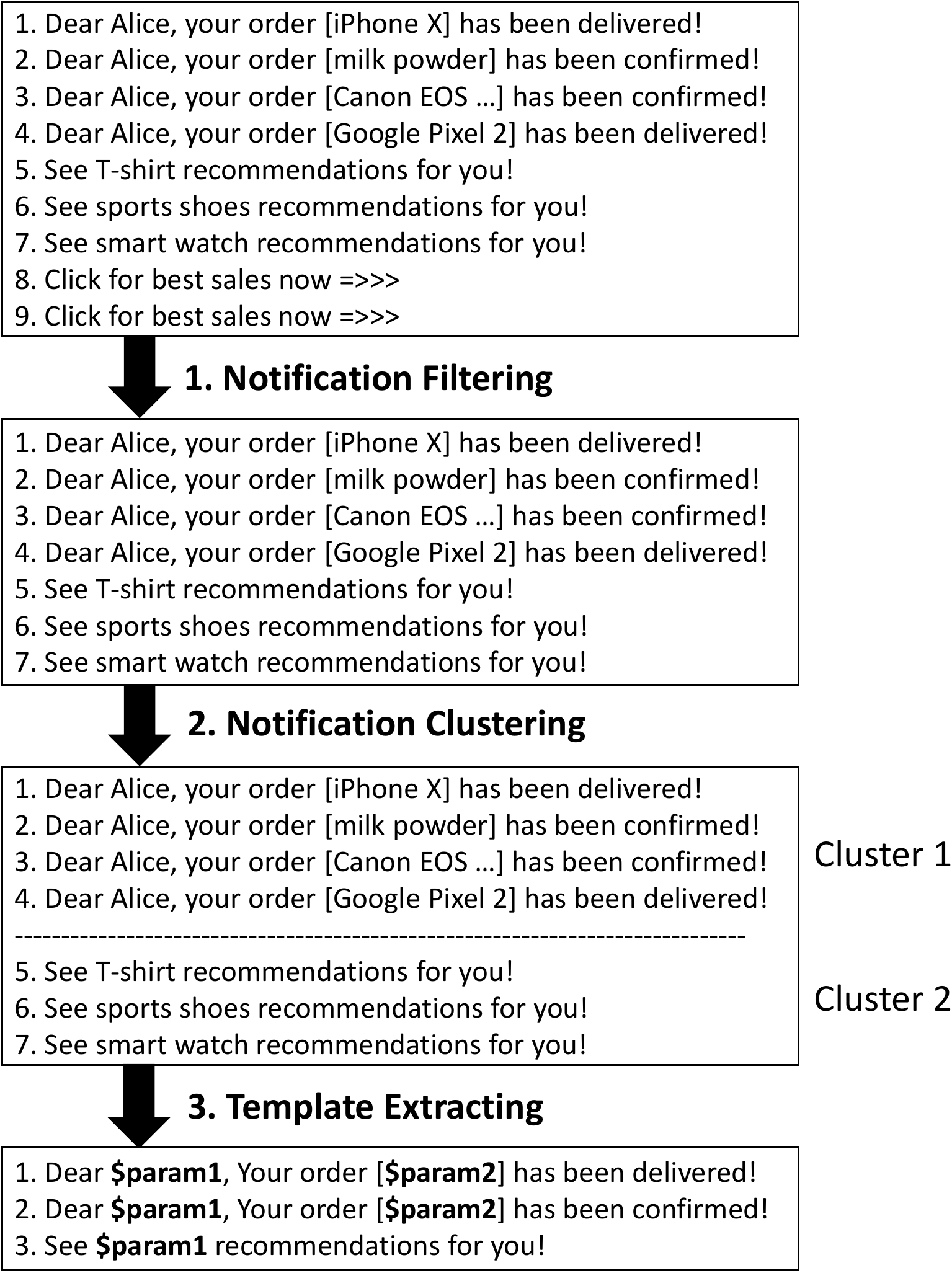}
\caption{An illustration of the template discovering process. Duplicated notifications and unstructured notifications are filtered in Step 1. Notifications generated with similar templates are clustered together in Step 2. Templates are extracted from each cluster in Step 3.}
\label{figure:template_discovering}
\end{figure}

In our solution, the template discovering process runs locally on each user's smartphone. It aims to identify templates that have at least two instances (i.e. two notifications generated from the same template) on a user device. The whole process involves several steps including notification filtering, notification clustering, and template extracting, as illustrated in Figure \ref{figure:template_discovering}.

Our insight is that the notifications generated from the same template are close to each other in terms of edit distance, while different from each other based on the parameters used.
Thus, it is possible to cluster the notifications such that notifications generated with the same template are grouped together. Then the template part and the parameter part of each notification can be differentiated by mining common patterns in each cluster.

\subsubsection{Notification Filtering}

We first preprocess the notifications by filtering out duplicated ones (such as system events, advertisements, etc.) and unstructured ones (such as text messages, emails, etc.). Such notifications usually do not contain personal knowledge and may introduce unnecessary computational load and inaccuracy to the clustering step.

There are two types of duplicated notifications, including local duplicates that appear multiple times on each user device and global duplicates that appear on multiple user devices.
Local duplicates are mainly repeated system event notifications such as ``Low battery'', ``Searching for GPS'', etc. They can be removed by simply comparing all local notifications.
Global duplicates are typically non-personalized promoting notifications sent to many users unchanged, such as ``Best sales!'', ``A new version is available.'', etc.
The global duplicates are identified by counting the total occurrences of each unique notification sentence among all users. To keep users' privacy, notifications are hashed before being uploaded to server for counting occurrences.
The commonly appeared hash values are then downloaded to user devices to help identify global duplicates.

Unstructured notifications, i.e. notifications generated without a template, are mainly messages or emails from other users. We identify and remove such notifications with several heuristics rules, for example checking the host app against a list of messenger apps and/or matching the notification text to known patterns, such as ``[NEW MAIL](.+)''.


\subsubsection{Notification Clustering}

After filtering, most of the remaining notifications are generated with templates. Given the fact that a template is a common subsequence of the notifications generated with it, extracting templates from notifications is similar to the task of longest common subsequence (LCS) mining.
However, mining LCS from these notifications can still be hard, as the notifications may be significantly different from each other.
A common solution, as used by Fu \textit{et al.} \cite{fu2009execution} for log analysis, is clustering the items before mining patterns from each cluster. Inspired by their work, we first cluster the notifications before extracting templates from them.

In order to group the notifications generated with the same template, we ought to select the correct clustering algorithm as well as the distance metric.
We notice that the notifications generated with the same template are quite close to each other in edit distance, where each edit operation can be adding, deleting, or replacing one word.
This is intuitive as the notifications are originally generated from templates by simple editing (adding entity values as parameters).
Meanwhile, the number of edits should be relatively small as compared to the length of the template. 

For example, in Figure \ref{figure:template_discovering}, the notification ``Dear Alice, Your order [iPhone X] has been delivered!'' is generated with the template ``Dear \texttt{\$param1}, your order [\texttt{\$param2}] has been delivered!''. In this example, the template occupies 10 words including punctuations, while the non-template part occupies only 3.
Thus we use the relative edit distance as the distance metric in our clustering algorithm:
$$edit\_distance(a, b) = \frac{\#\ minimum\ edits\ from\ a\ to\ b}{min(|a|, |b|)}$$
The clustering algorithm is easier to choose. As we do not know the exact number of templates, we use the DBSCAN algorithm \cite{ester1996dbscan} for clustering.

\begin{figure}[tbp]
\centering
\includegraphics[width=3in]{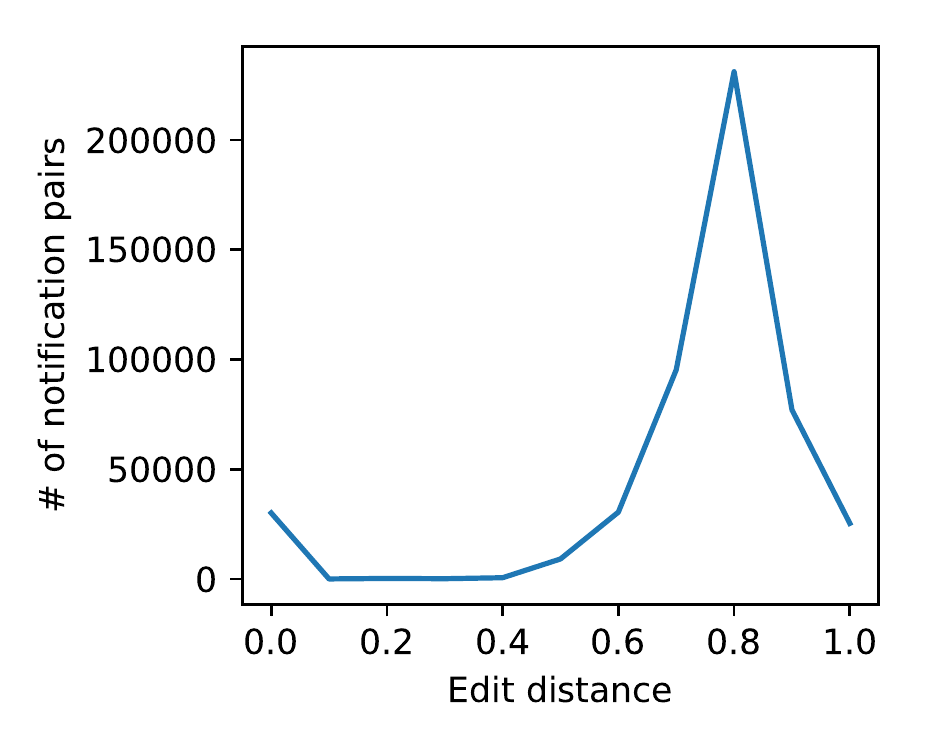}
\caption{The distribution of the edit distances of randomly selected 1,000 notifications. The threshold $\delta$ of the DBSCAN algorithm should be chosen from the flat region between the two peaks, in order to correctly cluster notifications.}
\label{figure:dbscan_k}
\end{figure}

According to the DBSCAN algorithm, items are grouped together only if the distance between any two items in the cluster is lower than a threshold $\delta$. Thus we need to determine the value of $\delta$ to run clustering.
If $\delta$ is selected too large, notifications generated with different templates will be grouped together (an extreme case is that all notifications are grouped into one cluster if $\delta = 1$), while if $\delta$ is too small, notifications generated with the same template may be separated into different groups.
To determine the proper threshold $\delta$, we investigate the distributions of edit distances between notifications in a few apps, like in \cite{fu2009execution}.
Specifically, we randomly picked 1000 notifications for each app in our dataset (the details of the dataset are described in Section \ref{ssec:dataset_overview}), and calculated edit distances for each pair of notifications. The distribution of the distances are then plotted out as a distribution graph.
There should exist two obvious peaks in the distribution graph. One is at a small distance value, meaning notifications generated from the same template have very short edit distances. Another one is at a large distance value, meaning most notification pairs that are not generated with the same template have long edit distances.
According to that, $\delta$ should be picked between these two peaks.
An example of the distribution graph is shown in Figure \ref{figure:dbscan_k}.
By repeating the process for multiple times and inspecting their distribution graphs, we choose $\delta=0.5$ in this paper.

Once the threshold is determined, we are able to cluster the notifications.
As shown in Figure~\ref{figure:template_discovering}, each group of notifications after clustering are generated with one template or several very similar templates.
This step also filters out a good deal of noisy data, i.e. notifications not belonging to any cluster.

\begin{figure}[tbp]
\centering
\includegraphics[width=3.5in]{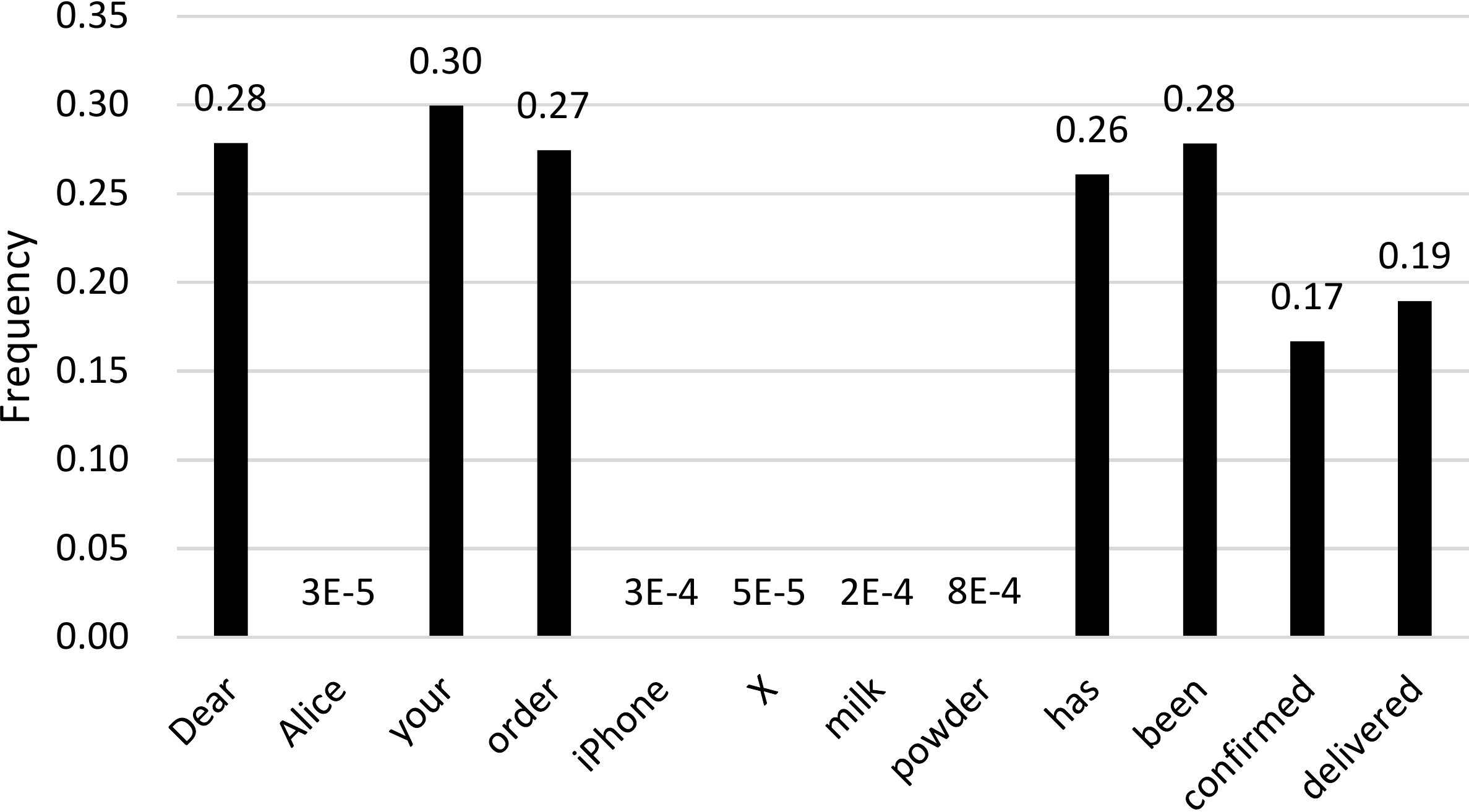}
\caption{Frequencies of the words in cluster 1 in Figure~\ref{figure:template_discovering} among all users. The parameter words ``Alice'', ``iPhone'', etc. are significantly less frequent than parameter words ``Your'', ``order'', etc.}
\label{figure:user_frequency}
\end{figure}

\subsubsection{Template Extracting}
\label{ssec:template_extracting}

As we mentioned before, we can identify potential templates by mining longest common subsequences (LCS) in each cluster of notifications, where the common sequences are  template bodies and the remaining part are left as parameter slots.
For example, by mining LCS from the cluster 1 in Figure \ref{figure:template_discovering}, we can get the template ``Dear Alice, your order \texttt{\$param1} has been \texttt{\$param2}''. This template has two parameter slots. ``iPhone X'' and ``milk powder'' are possible values to fill the first parameter slot \texttt{\$param1}, while ``delivered'' and ``confirmed'' for \texttt{\$param2}.

Unfortunately, there are two mistakes in the template discovered with LCS.
First, the user name ``Alice'', which should be a parameter in the template, is not correctly identified. This is because the user name is less variant across all notifications on the user device. Similar examples include the user's gender, profession, etc.
Second, ``delivered'' and ``confirmed'' are identified as parameter values, while they should not as they are not entities related to the user. The reason is that there are two similar templates in cluster 1 in Figure~\ref{figure:template_discovering}, while the LCS mining algorithm tries to extract only one template. 
The two mistakes are both due to parameter misidentification, i.e. parameters misidentified as template text or template text misidentified as parameter.

We make use of \emph{global word frequency} to address the problem. The global frequency of a word is defined as the number of users having at least one notification containing this word. The parameter values are usually more user-specific, thus having low global frequencies, while template words should have high global frequencies as they are usually user-agnostic.
Figure~\ref{figure:user_frequency} shows the global frequency of each word in cluster 1 in Figure~\ref{figure:template_discovering}.
The words ``Alice'', ``iPhone'', etc. have notably low global frequency as compared with non-parameter ones such as ``Dear'', ``order'', etc., thus they should be identified as parameter values.
The words ``delivered'' and ``confirmed'', although slightly less frequent than other template words, are identified as template text as they are still very common among all users. As a result, the template ``Dear Alice, your order \texttt{\$param1} has been \texttt{\$param2}'' extracted through LCS mining are corrected and split into two templates ``Dear \texttt{\$param1}, your order \texttt{\$param2} has been shipped'' and ``Dear \texttt{\$param1}, your order \texttt{\$param2} has been delivered''.

The above process is based on the \emph{global frequency} statistics, which requires a joint analysis of many user data on the server. 
To guarantee user privacy, we hash the words on user devices before uploading them to calculate the word frequency.
Specifically, each user uploads a list of hash values of the words used in the notifications of each app. The server calculates the global frequency of each word based on the uploaded hash values:
$$frequency(w) = \frac{\#\ users\ who\ uploaded\ hash(w)}{\#\ users\ who\ uploaded\ hashes}$$
The global frequencies are then downloaded to each user device to identify the parameters.

Finally, the notification templates extracted on user devices are uploaded to the server to determine the final set of templates.
We further process the templates on the server by filtering out incorrect ones based on several heuristic rules.
First, templates containing too many parameters and/or too few non-parameter words are unlikely to be correct. We exclude the templates with less than 5 non-parameter words or more than 3 parameters. Second, templates matching only a few users are excluded, as they are not general enough, or might contain personal information. Specifically, we excluded all templates matching fewer than 2 users.
We use the filtered set of templates for semantic analysis (as described in next sections) and knowledge extraction.

As templates used by each app are typically the same for different users, our approach only requires a small portion of users to discover templates on their devices and share the templates. Other users can directly download and use the templates without running the template discovering phase.
Meanwhile, sharing the templates should have little impact on privacy since the template itself does not contain any personal information.



\subsection{Template Semantic Rules}
\label{section:semantic_rules}


The discovered notification templates can be used to understand the sentence structure of the notifications.
To extract personal knowledge, we will need to further understand the semantics of each template.

Knowledge extraction in this kind of scenarios is typically modeled as as a slot filling problem \cite{Ji:2011:KBP}: given a document and a slot to fill (i.e. a knowledge triple with a pending entity), finding the boundaries of the slot filler (i.e. identifying the entity value).
The accuracy might be low if the document is poorly structured (F1 is around 35\% according to \cite{Ji:2011:KBP}).
Our approach can easily understand the sentence structure with the help of notification templates.
Thus the slot filling task is largely simplified: we do not need to determine the entity boundaries as they are automatically given by templates.

Due to the simplification, we are able to construct rules to extract knowledge from push notifications based on their templates.
We introduce \textit{template semantic rules} to help convert a notification to personal knowledge triples.
A template semantic rule is defined as a mapping from a notification template to a list of \textit{knowledge triple templates} (\textit{KTTs} in short). 

\begin{table}[tbp]
\centering
\caption{Examples of template semantic rules. The first column shows the examples of notification templates. The second column lists the templates of knowledge triples that can be extracted from the notification. Specifically, \texttt{u}, \texttt{\$p1} and \texttt{\$p2} are short for ``user'', ``parameter 1'' and ``parameter 2''.}
\label{table:semantic_rule_examples}
\begin{tabular}{p{1.8in}l}
\hline
\textbf{Notification template} & \textbf{Knowledge triple templates} \\ \hline
Good news! \texttt{\$p1} is on sale! & -              \\ \hline
Your flight to \texttt{\$p1} is delayed. & \texttt{<u, travelsTo, \$p1>} \\ \hline
\multirow{2}{*}{\shortstack[l]{Hi \texttt{\$p1}, your order \texttt{\$p2} has been\\ shipped.}} & \texttt{<u, name, \$p1>}  \\
& \texttt{<u, purchases, \$p2>} \\ \hline
\multirow{2}{*}{Mr. \texttt{\$p1}, please review the receipt.} & \texttt{<u, name, \$p1>}        \\
& \texttt{<u, gender, male>} \\ \hline
\multirow{2}{*}{\shortstack[l]{Here are some \texttt{\$p1} job\\ opportunities for you.}} & \texttt{<u, profession, \$p1>} \\
& \texttt{<u, status, job\_hunt>} \\ \hline
\end{tabular}
\end{table}


There are two types of \textit{KTTs} considered in our approach, including \textit{0-parameter KTTs} and \textit{1-parameter KTTs}.
An \textit{1-parameter KTT} can be used to generate different knowledge triples based on what parameter is used for the entity value. For example, \texttt{<user, travelsTo, \$param>} can use different location names (such as \texttt{New York City}, \texttt{China}, etc.) as the parameter. \texttt{<user, purchases, \$param>} can use different product names (such as \texttt{iPhone X}, \texttt{Nike Shoes}, etc.) as the parameter.
A \textit{0-parameter KTT} is a template with a fixed entity value, which can only generate one type knowledge triple. \textit{0-parameter KTTs} are suitable for attributive knowledge triples such as \texttt{<user, gender, male>}, \texttt{<user, status, job\_hunt>}, etc.


Table~\ref{table:semantic_rule_examples} shows some examples of template semantic rules.
For example, the non-personal template ``Good news! \texttt{\$param1} is on sale'' does not map to any personal knowledge triple.
``Hi \texttt{\$param1}, here are some \texttt{\$param2} job opportunities for you'' is a personal knowledge template that maps to three knowledge triple templates (\textit{KTTs}), including two \textit{1-parameter KTTs} (\texttt{<user, name, \$param1>} and \texttt{<user, profession, \$param2>}) and one \textit{0-parameter KTT} (\texttt{<user, status, job\_hunt>}).

With the template semantic rules, we are able to extract personal knowledge triples from the notifications formatted with the templates.
As shown in Figure~\ref{figure:overview}, we first identify the values of template parameters by matching the notifications with the templates. Then we fill the parameter values into the knowledge triple templates to obtain the actual knowledge triples.
For example, the notification ``Hi Alice, your order iPhone X has been shipped'' satisfies the template ``Hi \texttt{\$param1}, your order \texttt{\$param2} has been shipped'' with \texttt{\$param1} = ``Alice'' and \texttt{\$param2} = ``iPhone X''. According to the template's corresponding knowledge triple templates, we can extract two knowledge triples: \texttt{<user, name, Alice>} and \texttt{<user, purchases, iPhone X>}.

Once we have identified a template, we can then define its corresponding semantic rule either manually or through an automatic learning process. In an automated approach, we need to first manually label the semantic rules for a small set of notification templates. These manually labeled rules will be used in the automatic learning process to help infer the semantic meanings of newly found templates. The learned rules and manually labeled rules will both be used in our final step to extract the personal knowledge from new push notifications.


\subsection{Automated Template Semantic Rule Generation}

Although manually labeled semantic rules are the most accurate when used to understand notifications formatted with known templates, there might be a lot of unseen templates used by different or newer versions of apps. It is time-consuming and impractical to manually label semantic rules for all templates, as they may be generated by potentially millions of different apps.
Thus, we extend our system to automatically generate semantic rules for unseen templates, based on a set of manually defined semantic rules for known templates.

We model the problem as a sequence classification problem: given a notification template, predict what knowledge triple templates (\textit{KTTs}) it may have. Specifically, what \textit{0-parameter KTTs} the template has and what \textit{1-parameter KTTs} each template parameter belongs to.



\begin{figure*}[tbp]
\centering
\includegraphics[width=6.6in]{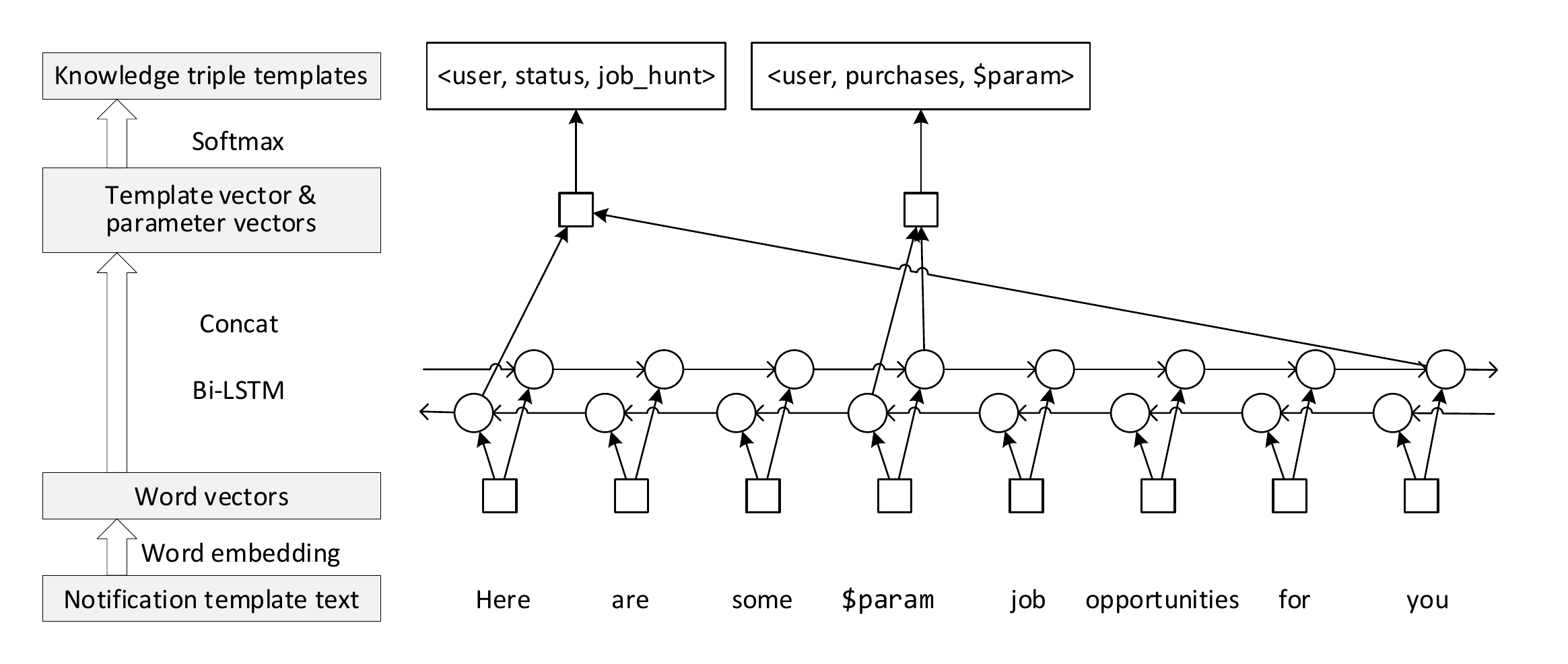}
\caption{The overview of our semantic rule prediction model. Given a notification template, we first represent each word and parameter in the template with word embedding. Then we compute a vector representation for the whole template and each parameter through a Bi-LSTM layer. Finally, the template vector is used for predicting \textit{0-parameter KTTs} and each parameter vector is used for predicting \textit{1-parameter KTTs}.}
\label{figure:lstm}
\end{figure*}

We use an RNN-based method to address the problem.
RNNs (Recurrent Neural Networks) are commonly used for NLP tasks such as PoS-tagging and named entity recognition, as they can effectively learn from the context of each word.
The context information is also very important for understanding the parameters in each template. For instance, given the template ``Your order \texttt{\$param1} has been shipped'', it is natural to guess that \texttt{\$param1} is the name of a user-purchased product based on its prefix ``Your order'', and ``has been shipped'' is also clearly describe the status of purchased product.
To make use of both the prefixes and suffixes of each parameter, we use the Bi-LSTM (Bidirectional Long Short-Term Memory) model to understand the template semantics.

Figure~\ref{figure:lstm} illustrates our model for automated semantic rule generation.
For each notification template, we first represent each word in the template as a vector through word embedding. We use an existing word embedding model pre-trained with fastText \cite{bojanowski2016enriching}, which is able to generate reasonable word embeddings for unseen words.
Reusing the pre-trained model enables us make use of the meanings of words learned from large corpus, thus facilitates training our model with relative small dataset.
The word vectors are then fed into a Bi-LSTM network, with which each word has a node capturing information from prefix words as well as a node capturing information from suffix words.
By concatenating the output of the two nodes for each parameter, we can generate a vector representation of the parameter.
Similarly, the whole template is represented as the concatenation of the outputs of the last word's forward node and the first word's backward node.
Finally, the parameter vector of each parameter is used to predict \texttt{1-parameter KTTs}, i.e. the knowledge triples that use the parameter as the entity value. The template vector is used to predict \texttt{0-parameter KTTs} whose entity values are fixed.

We use the manually labeled semantic rules (mappings from notification templates to knowledge triple templates) as to train the model. The hyperparameters are tuned based on several rounds of experiments. In the end, we use 200 hidden units for Bi-LSTM. We train the model for at most 15 epochs with the Adam optimizer, 1e-9 learning rate and 0.9 learning rate decay. Early stopping is adopted to prevent over-fitting.

The trained model is able to handle unseen notification templates.
Given a new template, the model can directly predict the knowledge triple templates it may have.
The mapping from a previously unseen template to the predicted knowledge triple templates is an automatically generated semantic rule.
Both the automatically-generated and manually-labeled semantic rules are used to extract knowledge triples from push notifications.

\section{Implementation and Evaluation}

We implemented two versions of the proposed knowledge extraction mechanism for production and experiments respectively:
\begin{enumerate}
\item The production version is implemented as an Android app. The app runs as a background service, collecting received notifications, identifying the template for each notification, and extracting personal knowledge from it.
The service also provides APIs for other apps to access the personal knowledge base. App developers can simply import our library and call \texttt{getPersonalKnowledge()} to get a stream of personal knowledge triples.
\item The experiment version is implemented entirely on a server. It takes a centralized dataset with all user notifications as the input and output the extracted knowledge triples for each user. However, we simulate the situation of the production version where the notifications are distributed on each user device, by processing each user data separately. This version of implementation is easier for conducting experiments as it does not require users to install our app.
\end{enumerate}

We evaluated our proposed approach by primarily looking at three aspects:
\begin{enumerate}
\item Can our system discover new personal knowledge templates from smartphone notifications accurately?
\item Can our system understand the meaning of the templates accurately, especially for previously unseen templates? Specifically, what is the accuracy of the template semantic analysis?
\item Can our system run on real user devices without introducing too much overhead?
\end{enumerate}
To answer these questions, we conducted experiments on a dataset of push notifications from real users.

\subsection{Dataset Overview}\label{ssec:dataset_overview}

\begin{figure*}[tbp]
    \centering
    \begin{subfigure}[b]{0.3\textwidth}
        \includegraphics[width=\textwidth]{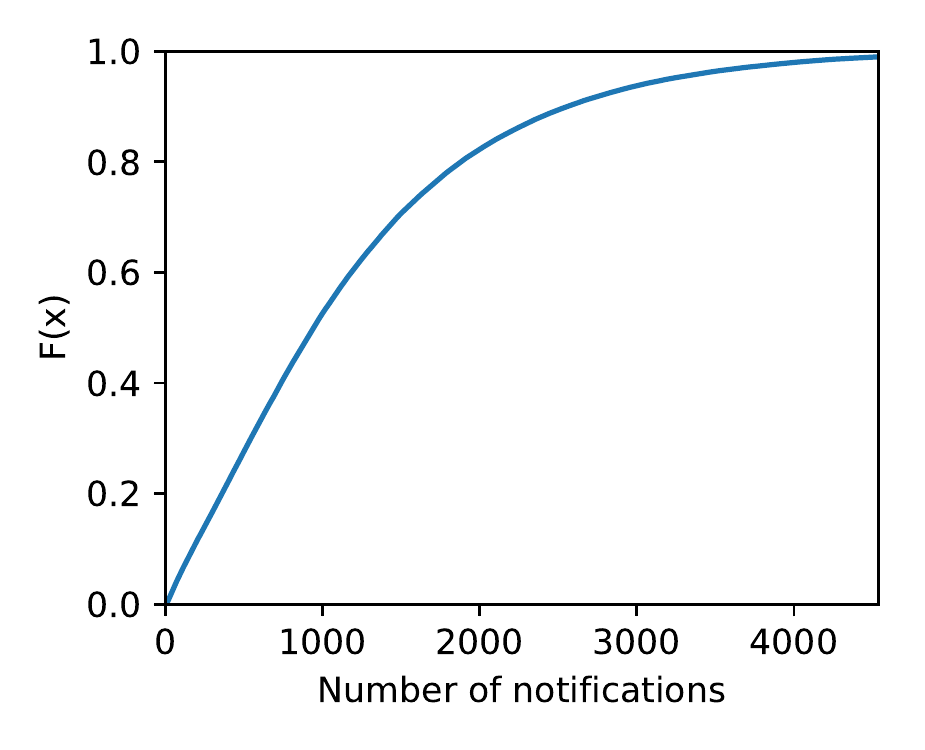}
        \caption{The number of notifications per user.}
        \label{figure:cdf:num_notis}
    \end{subfigure}
    \hspace{0.5cm}
    \begin{subfigure}[b]{0.3\textwidth}
        \includegraphics[width=\textwidth]{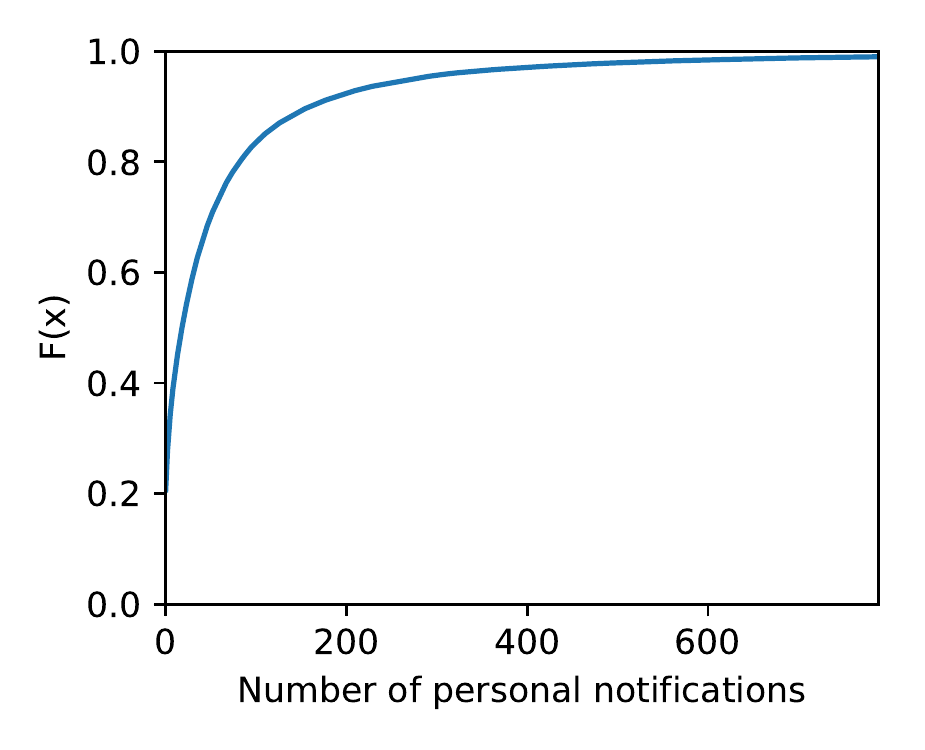}
        \caption{The number of personal notifications per user.}
        \label{figure:cdf:num_personal_notis}
    \end{subfigure}
    \hspace{0.5cm}
    \begin{subfigure}[b]{0.3\textwidth}
        \includegraphics[width=\textwidth]{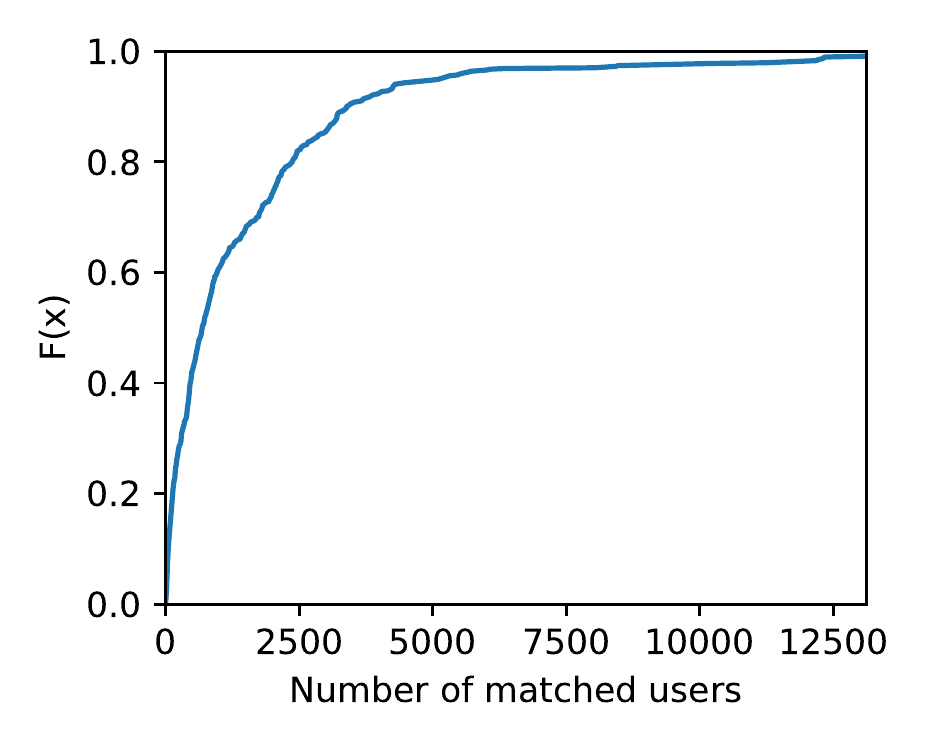}
        \caption{The number of matched users per template.}
        \label{figure:cdf:num_matched_users}
    \end{subfigure}
    \caption{Statistics of our dataset, including the distributions (CDFs) of the number of notifications per user, the number of personal notifications per user, and the number of matched users per personal knowledge template.}
    \label{figure:cdfs}
\end{figure*}

The dataset we used in the experiments contains 119,289,901 notifications from 100,000 smartphone users, obtained through a mobile service provider in China. As these notifications may contain sensitive user information, all data have been collected from a group of designated test users in accordance with the policies of the service provider. We have strictly followed the ``terms and conditions'' specified by the smartphone provider with respect to these test users in our study. For example, the identities of all users have been anonymized, while all push notifications have been kept on the servers within the provider's company throughout the whole process. 

The notifications were generated by 2,658 apps over 30 days from March to April 2018. Each notification entry is consisted of:
\begin{itemize} 
\item A user ID: the unique identity of the sampled smartphone. Each user ID is anonymized for security and privacy reasons.
\item An app ID: the unique identity of the app that the notification belongs to.
\item A timestamp: the time when the notification was pushed to the user's smartphone.
\item Notification content: the notification title followed by the text body. All numbers in the notifications are elided for security and privacy reasons.
\end{itemize}

Figure~\ref{figure:cdf:num_notis} shows the distribution of the number of notifications per user in our dataset. Most users (around 50\%) have more than 1000 notifications and about 20\% of the users have more than 2000 notifications.
On average, there are about 40 notifications for each user per day in our dataset, which is a subset of users' notifications. The reason is that we have only obtained the notifications that are pushed through a certain service provider, instead of all the notifications on a smartphone. However, while there is some bias in our data set, we believe that our technique can still generalize.

\subsection{Accuracy of Notification Template Discovering}\label{ssec:temp_dis_accuracy}

We simulated the scenario that the notifications are distributed on users' smartphones, and used the template discovering method described in Section~\ref{section:template_discovering} to identify notification templates. 

In total, we discovered 2,788 templates belonging to 409 apps from the dataset.
We manually labeled the discovered templates, determining whether each template contains personal knowledge and whether it is correct (by correct we mean that the template correctly identifies the boundaries of parameters). The result shows that there are 2,163 personal knowledge templates, among which 2,006 (92.7\%) are correct, and 625 non-personal templates, among which 414 (66.2\%) are correct. The overall correctness ratio of template discovering is 86.8\%.
The correctness for non-personal templates is relatively low because those notifications are usually less-structured. Such notifications include top news, sales information, etc., many of which are manually crafted without a template. However, it is not a huge problem as we are not going to extract knowledge from these non-personal templates anyway.
Figure~\ref{figure:cdf:num_matched_users} shows the distribution of the number of matched users per personal knowledge template (i.e. the users who have one or more notifications matching the template). On average, each personal knowledge template has matched 1,571 users.

\begin{table*}[tbp]
\centering
\caption{Examples of the notification templates discovered from our dataset. The \#users column shows the number of users that have notifications matching the template. The \#notifications column shows the total number of notifications matching the template. Note that the notification templates are translated from Chinese.}
\label{table:template_examples}
\begin{tabular}{cp{3.8in}rr}
\hline
\textbf{Category}                    & \textbf{Notification template examples}                                      & \textbf{\#users} & \textbf{\#notifications} \\ \hline
\multirow{3}{*}{User profile}        & - \texttt{\$param1} just downloaded your resume, talk with them now! & 187                       & 1,003                             \\
                                     & - Dear \texttt{\$param1}, see new comments from your friends.                          & 497                       & 2,021                             \\
                                     & - Now hiring! See \texttt{\$param1} job opportunities near \texttt{\$param2}.     & 1,861                       & 20,533                            \\ \hline
\multirow{3}{*}{Social} & - Your friend \texttt{\$param1} sent you a message.              & 328                      & 21,366                            \\
                                     & - \texttt{\$param1}: \texttt{\$param2} is now live streaming.                            & 3,385                       & 47,904                            \\
                                     & - Hi \texttt{\$param1}, see what your friend \texttt{\$param2} is talking about you!      & 824                       & 3,958                             \\ \hline
\multirow{3}{*}{Location} & - [Weather Forecast] There will be rainy near \texttt{\$param1} this afternoon.              & 1,424                       & 3,942                            \\
                                     & - Warning: Traffic is heavy near \texttt{\$param1} ...                            & 1,262                       & 2,284                            \\
                                     & - These delicious restaurants near \texttt{\$param1} are popular! & 1,474                       & 1,852                             \\ \hline
\multirow{3}{*}{Shopping}     & - Your order \texttt{\$param1} is on its way.                                   & 25,100                     & 118,757                           \\
                                     & - Shipping notice: The item \texttt{\$param1} has been shipped!                 & 13,779                       & 31,695                            \\
                                     & - Here are some good matches with the \texttt{\$param1} that you purchased.     & 6,064                       & 6,760                             \\ \hline
\end{tabular}
\end{table*}

Table~\ref{table:template_examples} shows some examples of the personal knowledge templates discovered from our dataset. We selected three most common templates (i.e. the templates that match most users) in each of the three knowledge categories.
As we can see, a notification can contain multiple categories of personal knowledge. For example, the template ``Hi \texttt{\$param1}, see what your friend \texttt{\$param2} is talking about you!'' contains both user profile information (the user's name) and social relationship information (name of the user's friend).
It is also interesting to see that there are other categories of personal knowledge except for the three categories we considered. For example, ``Now hiring! See \texttt{\$param1} job opportunities near \texttt{\$param2}.'' contains the user's job preference information, which is also an important type of personal knowledge.

In total, we found that about 5.7\% of all notifications contain some kind of personal information\footnote{The proportion can be higher if we consider more types of personal knowledge beyond what have been defined in this paper.}. Specifically, 6,824,002 out of 119,289,901 notifications are identified as personal notifications, as they can match one of the discovered personal knowledge templates.
The distribution of the number of personal notifications per user is shown in Figure~\ref{figure:cdf:num_personal_notis}.
Around 10\% of the users have more than 200 personal notifications.
On average, there were at least 68 push notifications for each user that contain personal knowledge.
As we only considered a subset of notifications due to the limitation of our dataset, we believe that more personal notifications can be found on actual smartphones.

\subsection{Accuracy of Template Semantic Analysis}

We also conducted experiments to evaluate how well our system can correctly understand the semantics of previously unseen templates.
We used the 2,788 notification templates from 409 apps discovered with our template discovering method, as described in Section \ref{ssec:temp_dis_accuracy}.

\begin{figure}[tbp]
\centering
\includegraphics[width=3.5in]{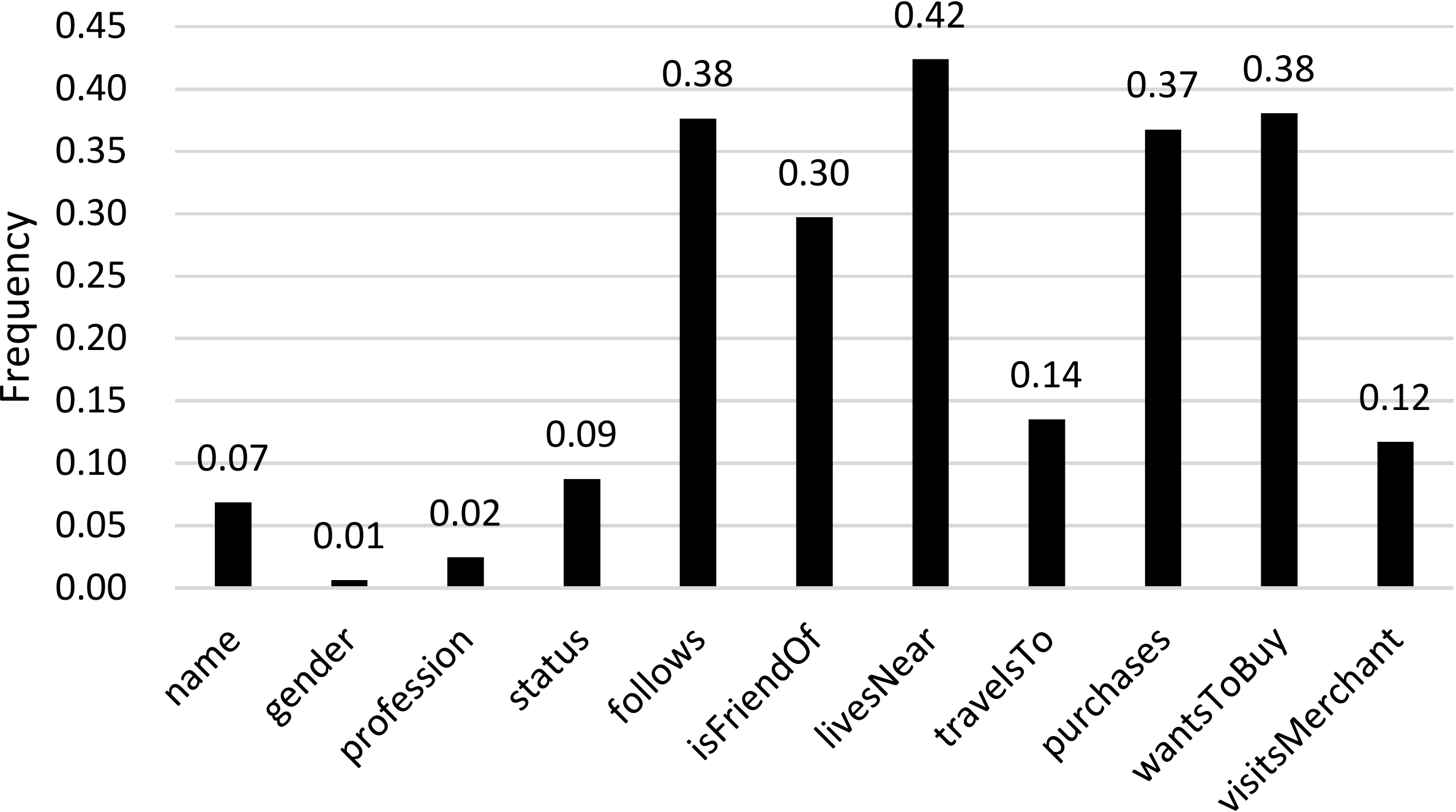}
\caption{The number of users per knowledge type.}
\label{figure:knowlege_types}
\end{figure}

We manually labeled the knowledge triple templates for each notification template according to our personal knowledge ontology (Table~\ref{table:ontology}) as the ground truth.
Figure~\ref{figure:knowlege_types} shows the number of users that have each type of knowledge (a user is identified to have a type of knowledge if one of his/her notifications matches a template in that knowledge type).
As we can see, only a small portion of users have user-profile-related knowledge in their notifications, while lots of them can be extracted knowledge in social relationship, location, and shopping categories.
This is because only few apps put user's personal information (name, gender, profession, etc.) into notifications and even fewer users are using these apps.

\begin{table*}[tbp]
\centering
\caption{Accuracy of template semantic analysis. We used our model to predict labels (knowledge triple templates) for unseen notification templates, including the templates used by unseen apps and the new templates of existing apps. 
}
\label{table:semantic_accuracy}
\begin{tabular}{ccrrR{2cm}rR{2.7cm}r}
\hline
\multirow{2}{*}{\textbf{KTT Category}}     & \multirow{2}{*}{\textbf{KTT}} & \multirow{2}{*}{\textbf{\#apps}} & \multirow{2}{*}{\textbf{\#templates}} & \multicolumn{2}{c}{\textbf{Templates of unseen apps}} & \multicolumn{2}{c}{\textbf{New templates of existing apps}} \\
                              &             &         &          & \textbf{precision}    & \textbf{recall}       & \textbf{precision}    & \textbf{recall} \\ \hline
\multirow{2}{*}{0-param}      & \texttt{<u, status, job\_hunt>}& 8  & 48  & 92.31\% & 71.79\% & 94.13\% & 93.89\% \\
                              & \texttt{<u, status, car\_hunt>}& 10 & 97  & 92.40\% & 77.37\% & 91.27\% & 89.68\% \\ \hline
\multirow{7}{*}{1-param}      & \texttt{<u, name, \$p>}    & 22 & 90  & 87.81\% & 85.45\% & 93.77\% & 95.91\% \\ 
                              & \texttt{<u, follows, \$p>}    & 45 & 217 & 90.37\% & 76.21\% & 90.85\% & 93.89\% \\
                              & \texttt{<u, isFriendOf, \$p>}  & 129& 485 & 87.12\% & 86.52\% & 90.21\% & 92.95\% \\
                              & \texttt{<u, livesNear, \$p>}   & 16 & 157 & 74.13\% & 71.40\% & 93.94\% & 87.58\% \\
                              & \texttt{<u, travelsTo, \$p>}   & 9  & 39  & 93.38\% & 86.72\% & 89.66\% & 92.10\% \\
                              & \texttt{<u, purchases, \$p>}   & 22 & 83  & 73.30\% & 86.85\% & 87.50\% & 91.79\% \\
                              & \texttt{<u, wantsToBuy, \$p>}  & 41 & 234 & 76.93\% & 84.45\% & 82.31\% & 90.88\% \\ \hline
\multirow{1}{*}{Overall} & & 302& 1450& 84.50\% & 81.86\% & 89.69\% & 92.08\% \\ \hline
\end{tabular}
\end{table*}

Each knowledge triple template is a label in our classification model, and each notification template can have zero or more labels.
Labels are selected in order that there are enough samples (notification templates having the label) for machine learning.
For example, the knowledge triple templates with \texttt{gender} relation and \texttt{profession} relation are not used as label in this experiment as they.
In total, we selected two \texttt{0-parameter KTTs} to evaluate the classification of templates and seven \texttt{1-parameter KTTs} to evaluate the classification of parameters, as shown in Table~\ref{table:semantic_accuracy}.


We considered two situations where we need to predict semantic rules for unseen notification templates. The first is that when a new app is added to our system, all of the notification templates used by that app are unseen. The second is that when an existing app is updated, it may use a new template to bring information to its users.
We designed two experiments to evaluate our system's performance for both situations:
\begin{itemize}
\item We first ran a 5-fold cross-validation to check whether our system is able to handle unseen templates from unseen apps.
For each label, we randomly divided the 417 apps into 5 sets such that each set had approximately equal amount of apps whose templates have the label. For example, each set will have about 9 apps that contain \texttt{<u, follows, \$p>} knowledge triples, as there are in total 45 apps containing such knowledge.
In each fold, we used 4 sets of templates for training and the remaining set of templates for predicting.
\item In the second experiment, we also use 5-fold cross-validation, but with different partitioning method. For each app, we randomly divided its templates to 5 sets. We used 4 sets for training and the remaining 1 set for predicting in each fold.
\end{itemize}

The results of the both experiments are shown in Table~\ref{table:semantic_accuracy}. Overall, our semantic model is able to accurately predict labels (i.e. generate semantic rules) for unseen templates.
The accuracy of predicting labels for new templates of existing apps is high (89.41\% precision and 92.19\% recall), which is not a surprise because the new templates of an app are usually similar to its old templates.

For unseen apps' templates, the precision (83.62\%) and the recall (82.56\%) are both lower than the other situation. Is is because that different apps may use significantly different ways to express same types of knowledge.
For example, a live streaming app (such as Twitch) may notify users about the updates of their subscribed anchors using ``\texttt{\$param} is live streaming.'', while an online publishing app (such as Medium) may notify users about the updates of their favorite authors using ``\texttt{\$param} posted a new article.'', both notification templates express a \emph{following} social relationship while using totally different vocabularies.
However, the accuracy is acceptable for most common use cases of personal knowledge, such as recommender systems and conversational bots.

Among the knowledge relations considered in our evaluation, the accuracy for ``livesNear'', ``purchases'' and ``wantsToBuy'' is relatively low.
One major reason is that these relations can be expressed in a wide range of ways, while our dataset only contains a limited number of apps, each using a specific way to express the knowledge.
We think this problem can be tempered by adding more training data, such as more labeled notification templates from other apps.

\subsection{Client Overhead}

We also evaluated the overhead that our approach may bring to end-users by running the system on a real device.
The device we used in this experiment is a Nexus 5 phone running Android 7.1.2.
Most of the time our app simply log the received push notifications and save them to a local database, which only consumes little storage and computational resource.
Occasionally (e.g. once a week), our app need to run the knowledge extraction phase. 
According to our experiment, extracting knowledge triples from 10,000 notifications only took around 5.8 seconds.

Some users will also need to run the template discovering phase in order to contribute knowledge templates to the server.
We experimented with different amount of notifications and found that it took about 2.8, 38.7, and 770.8 seconds to process (including clustering notifications, extracting templates from the clusters, and uploading the discovered templates) 100, 1,000, and 10,000 notifications respectively.
As our system only requires a small portion of users to run the template discovering phase, and both the template discovering phase and the knowledge extraction phase only need to be performed occasionally, we believe it is little overhead for most normal users.

\section{Limitations and Future Work}

In this section, we highlight some of the limitations of our system and discuss possible solutions.

\textbf{Strong privacy guarantee.} Our system is privacy-preserving because it only uploads the notification templates to the server. However, it is not a strong privacy guarantee because the uploaded templates may contain personal information if the templates are incorrect. One possible solution is to scan potential templates for sensitive information before uploading, e.g. using Named Entity Recognition techniques.

\textbf{Real-world scenario.} Our system is evaluated with a dataset provided by a push notification service provider, which only contains remotely-generated notifications from a small subset of apps. The real scenario might be different as we will be able to access all notifications on users' smartphones. In the future, we would like to deploy our system and evaluate the performance of our system in the real-world scenario.

\textbf{Comprehensive personal knowledge ontology.} We considered three common categories of personal knowledge in our implementation. However, a lot of other knowledge categories can be found in push notifications, such as work information, travel information, etc. Meanwhile, our knowledge ontology is specifically designed for personal knowledge in push notifications. There ought to be a more complete and formal ontology of personal knowledge, like the one defined in Schema.org \cite{schema.org} for world's knowledge.

\textbf{Other ways to obtain notification templates.} Our approach requires a portion of users to upload discovered notification templates for offline learning. This requirement might be hard to fulfill if our system does not have enough users. We can solve this problem by using other methods to obtain an initial set of notification templates, such as extracting the templates from application code through static analysis, or generating notifications by automatically testing the apps \cite{li:icse2017:droidbot}.

\section{Related Work}

\subsection{Knowledge Base Population}

Automated knowledge base population is an important problem in both academia and industry. There are mainly four groups of approaches in this area according to \cite{knowledgeVault}: 
YAGO \cite{yago}, YAGO2 \cite{yago2}, DBpedia \cite{dbpedia}, and Freebase \cite{freebase} extracts information from structured data sources such as Wikipedia infoboxes; 
Reverb \cite{reverb}, OLLIE \cite{ollie}, PROSMATIC \cite{prismatic} use open information (schema-less) extraction techniques applied to the entire web;
NELL \cite{nell}, PROSPERA \cite{PROSPERA} and Elementary \cite{elementary} extract information from the entire web, bus use a fixed ontology;
Approaches like Probase \cite{probase} construct taxonomies (is-a hierarchies), as opposed to general knowledge bases with multiple types of predicates.
Unlike these approaches that try to construct knowledge bases in the public domain, we focus on personal knowledge related to each individual.

Extracting knowledge from natural language text is typically modeled as a slot filling task \cite{Ji:2011:KBP}, i.e. adding information about an entity to the knowledge base based on the language semantics. The state-of-the-art approaches \cite{zhang2016joint,yao2014recurrent} framed the task as sequence labeling and applied neural networks to predicting the boundaries and labels of entity values.
However, the data source we used to extract knowledge from is mobile notification text, which is significantly different from the data sources of the world's knowledge, as notifications are mainly structured with templates.
Our knowledge extraction method is tailored for such kind of data based on template learning and template understanding.

\subsection{Personal Knowledge Extraction}

Personal knowledge is the basis to many personalized services, thus how to collect personal knowledge has attracted many researchers' and companies' interests. The commonly used data sources for extracting personal knowledge include search queries, conversational dialogs, SMS messages, sensor, and UI content.
For example, 
various approaches \cite{TMC:Vhaduri:places,TMC:Do:Places,TMC:Samaan:Mobility} have been proposed to mine users' places of interest and mobility patterns from mobile data.
Min \textit{et al.} \cite{use_case6:cscw2013:social_relationships} and Cruz \textit{et al.} \cite{TMC:Cruz:Contacts} analyzed communication logs and wireless access logs respectively to infer users' social relationships.
Spolaor \textit{et al.} \cite{TMC:Spolaor:DELTA} presented a data extraction tool named DELTA, which can extract UI interaction data for user habit analysis.
Li \textit{et al.} \cite{personalKnowledgeUtterances} introduce a statistical language understanding approach to automatically construct personal knowledge graph from conversational dialogs.
Google Now on Tap and Bing Snapp can identify entities in the UI content displayed by apps and show information related to these entities.

As the data sources of personal knowledge are often privacy-sensitive, a lot of approaches have been proposed to mitigate the privacy concern during knowledge mining. The most commonly-used method is to do the computation locally on users' devices. For example, Shen \textit{et al.} propose a client-side web search agent \cite{UCAIR} to build user models for search personalization. PrivacyStreams \cite{privacystreams} introduce an Android library for app developers to process personal data locally and transparently. Appstract \cite{appstract:MobiCom2016} use an offline user-agnostic learning phase and an on-device predicting phase to preserve the privacy of UI content analysis.
Inspired by Appstract, our system also adopts a two-phase method to minimize users' privacy concern.

\subsection{Notification Analysis}

Prior to our work, push notifications have been studied by researchers from different aspects. 
Most researchers have focused on the disruptive nature of push notifications \cite{notifications:CHI2014,MyPhoneAndMe,noti_in_situ_study,noti_prefminer} in order to guide the design implementation of better notification systems.
For example, Shirazi \textit{et al.} \cite{notifications:CHI2014} analyzed a large scale of notifications and revealed the differences in the importance of notifications. Mehrotra \textit{et al.} \cite{MyPhoneAndMe,noti_prefminer} analyzed how notifications with different content, sender and context can cause disruption to users.
Other researchers have also explored the effects of push notifications to foster meta-learning \cite{noti_learning} and self-logging \cite{noti_logging}. 
Our work does not aim to improve the notification mechanisms or use notifications to influence users' behaviors, instead, we make use of notification content for a broader purpose: building a personal knowledge base. The knowledge base can be used to support all kinds of personalized services.

\subsection{Wrapper Generation}

Wrapper in data mining is a program that extracts content of a particular information source and translates it into a relational form.
The aim of a wrapper is to locate relevant information in semi-structured data and to put it into a self-described representation for further processing \cite{wrapper_induction}.
Typically, wrappers are used to extract structured data, such as telephone directories, product catalogs, etc. from web pages formatted with fixed HTML templates.
Wrappers can be generated manually by experts, semi-automatically through supervised learning \cite{wrapper_induction,hierarchical_wrapper,lixto}, or automatically through unsupervised learning \cite{crescenzi2001roadrunner,chang2003automatic,fu2009execution}.
Our work deals with another form of wrapper: push notification templates.
We used an unsupervised approach to discover the templates and supervised machine learning to understand the semantic rules of the templates.

\section{Concluding Remarks}

This paper proposes an approach for extracting personal knowledge from smartphone push notifications.
It is able to automatically identify templates from notification text using pattern mining techniques, and then understand the semantics of the templates through supervised machine learning.
The approach is privacy-preserving as only the templates might be uploaded to the server for labeling and learning.

We have implemented a prototype system on Android and evaluate it with three common categories of personal knowledge. Experiments on about 120 million push notifications from 100,000 smartphone users show that we are able to discover and understand notification templates accurately, while successfully use the templates to extract personal knowledge from push notifications.


\bibliographystyle{IEEEtran}
\bibliography{IEEEabrv,citation}

\end{document}